\documentclass[aps,prd,groupedaddress,showpacs,twocolumn]{revtex4}
\usepackage{graphicx}
\usepackage{dcolumn}
\usepackage{bm}
\begin{document}

\title{Reaction $e^+e^-\to\pi^+\pi^-\pi^+\pi^-$ at energies $\sqrt{s}\leq$ 1 GeV.}

\author{N.~N.~Achasov}
\email[]{achasov@math.nsc.ru} \affiliation{Laboratory of
Theoretical Physics, S.~L.~Sobolev Institute for Mathematics,
630090, Novosibirsk, Russian Federation
}%
\author{A.~A.~Kozhevnikov}
\email[]{kozhev@math.nsc.ru} \affiliation{Laboratory of
Theoretical Physics, S.~L.~Sobolev Institute for Mathematics,
630090, Novosibirsk, Russian Federation}

\date{\today}
\begin{abstract}
The cross section of reaction $e^+e^-\to\pi^+\pi^-\pi^+\pi^-$ is
calculated for energies $0.65\leq \sqrt{s}\leq1$ GeV in the
framework of the generalized hidden local symmetry model. The
calculations are compared with the  data of CMD-2 and BaBaR. It is
shown that the inclusion of heavy isovector resonances
$\rho(1450)$ and $\rho(1700)$ is necessary for reconciling
calculations with the data. It is found that at $\sqrt{s}\approx1$
GeV the contributions of  above resonances are much larger, by the
factor of 30, than the $\rho(770)$ one, and are amount to a
considerable fraction $\sim0.3-0.6$ of the latter at $\sqrt{s}\sim
m_\rho$.
\end{abstract}
\pacs{11.30.Rd;12.39.Fe;13.30.Eg}

\maketitle

Among  chiral models aimed at the description of interactions of
the pseudoscalar mesons  with the low lying vector and axial
vector ones, see the review \cite{meissner88} and references
therein, the most elegant  is the generalized hidden local
symmetry (GHLS) model \cite{bando88}. It relates all coupling
constants to only the pion decay constant $f_\pi$ and
$g_{\rho\pi\pi}$, and  accounts for anomalous processes in a way
that does not break low energy theorems.  Strikingly, but this
very popular model was not scrutinized in the processes with
sufficiently soft pions where one can rely on the tree
approximation. The purpose of the present paper is to fill this
gap by plotting the $e^+e^-\to\pi^+\pi^-\pi^+\pi^-$ reaction cross
section in the GHLS model and comparing the results with available
data CMD-2 \cite{cmd2} and BaBaR \cite{babar}. When so doing, we
use our recent calculations of the $\rho\to4\pi$ decay amplitudes
\cite{ach00b,ach05} to account for the resonant production
$e^+e^-\to\rho\to\pi^+\pi^-\pi^+\pi^-$.  Note that excitations
curves in \cite{ach00b} do not include the $a_1\pi$ intermediate
state \cite{ach05} nor the contact non-resonant contributions
$e^+e^-\to\gamma^\ast\to\rho\pi\pi\to4\pi$,
$e^+e^-\to\gamma^\ast\to a_1\pi\to4\pi$ whose explicit form is
found here.

The ingredients for  the  amplitude with the resonant $\rho$ meson
are given in \cite{ach00b,ach05}. The Lagrangian of the direct
photon coupling  is
\begin{eqnarray}
{\cal L}_{\rm photon}&=&-e{\cal A}_\mu\left(2gf^2_\pi\rho^0_\mu-
\frac{\pi^+\pi^-}{2f^2_\pi}[{\bm\pi}\times\partial_\mu{\bm\pi}]_3
-\right.\nonumber\\&&\left.2g\rho^0_\mu\pi^+\pi^-+2gf_\pi[{\bm\pi}\times{\bm
a}_\mu]_3\right),\label{photon}\end{eqnarray}where
$g=g_{\rho\pi\pi}$, and ${\cal A}_\mu$, ${\bm a}_\mu$, ${\bm\pi}$
stand for the photon four-vector potential, $a_1(1260)$, $\pi$
meson field, respectively. Boldface characters refer to isotopic
vectors. Given are only the terms necessary for the
$\pi^+\pi^-\pi^+\pi^-$ final state, and the contributions of the
second order in electric charge $e$ are neglected. Note that the
contact $\gamma^\ast\to\pi^+\pi^-$ and
$\gamma^\ast\to\pi^+\pi^-\pi^+\pi^-$ vertices cannot be
simultaneously eliminated in HLS, while   the contact
$\gamma^\ast\to\pi^+\pi^-$ vertex is eliminated in HLS by the
parameter choice \cite{bando88}.

It is suitable to represent the energy dependence of the
$e^+e^-\to\pi^+\pi^-\pi^+\pi^-$ reaction cross section in the form
\begin{equation}
\sigma_{e^+e^-\to4\pi}(s)=\frac{12\pi m^3_\rho\Gamma_{\rho
e^+e^-}(m_\rho) \Gamma^{\rm
eff}_{\rho\to4\pi}(s)}{s^{3/2}|D_\rho(q)|^2},
\label{curve}\end{equation}where the  leptonic width of the vector
meson $V$ on the mass shell looks as
\begin{equation}
\Gamma_{Ve^+e^-}(m_V)=\frac{4\pi\alpha^2m_V}{3f^2_V},\label{leptwidth}
\end{equation} and  $s=q^2$ is the total energy
squared in the center-of-mass system. The function   $\Gamma^{\rm
eff}_{\rho\to4\pi}(s)$  in (\ref{curve}) is evaluated  with the
effective $\rho\to4\pi$ decay amplitude $M^{\rm
eff}_{\rho\to4\pi}\equiv M^{\rm
eff}_{\rho_q\to\pi^+_{q_1}\pi^+_{q_2}\pi^-_{q_3}\pi^-_{q_4}}$
which includes both the resonant contribution
$e^+e^-\to\gamma^\ast\to\rho\to\pi^+\pi^-\pi^+\pi^-$  and  the
contact one $e^+e^-\to\gamma^\ast\to\pi^+\pi^-\pi^+\pi^-$.  In the
lowest order in electromagnetic coupling constant this amplitude
is given by the expression
\begin{equation}
M^{\rm eff}_{\rho\to4\pi}=
\frac{g_{\rho\pi\pi}}{f^2_\pi}\epsilon_\mu(A_1q_{1\mu}+A_2q_{2\mu}+A_3q_{3\mu}+A_4q_{4\mu}),
\label{meff}\end{equation}where $\epsilon_\mu$ stands for the
polarization four-vector of the virtual $\rho$ meson, and
$A_a\equiv A_a(q_1,q_2,q_3,q_4)$, $a=1,2,3,4$ are dimensionless
invariant functions. $A_1=-1+(1+\widehat{P}_{34})B_1$, where
\begin{eqnarray}
B_1&=&\frac{2}{D_\pi(q-q_1)}\left[\frac{m^2_\rho}{D_{\rho23}}(q_4,q_2-q_3)-(q_2,q_3)\right]
-\nonumber\\&&D_\rho(q)\left(\frac{1}{D_{\rho14}}-\frac{1}{2m^2_\rho}\right)-
\frac{(1-\widehat{P}_{23})}{4D_{a_1}(q-q_1)}\times\nonumber\\&&
\left\{\frac{1}{D_{\rho23}}\left[4(q_2,q_4)(2q-q_1,q_3)-\right.\right.\nonumber
\\&&\left.\left.(2q-q_1,q_2)(q_4,q-q_1)+(2q-q_1,q_4)(q_2,q_4)\right]
-\right.\nonumber\\&&\left.\frac{1}{2m^2_\rho}(q_2,2q-2q_1+q_4)(2q-q_1,q_3)-
\right.\nonumber\\&&\left.\frac{4(q_2,q_4)(q_2+q_3)^2}{m^2_{a_1}}\left[q^2+D_\rho(q)\right]\left(\frac{1}{D_{\rho23}}-
\right.\right.\nonumber\\&&\left.\left.
\frac{1}{8m^2_\rho}\right)\right\}-\frac{3(q,q_2)-m^2_\pi-D_\rho(q)}{4D_{a_1}(q-q_2)}
\times\nonumber\\&&\left[\frac{(q_4,4q_3-q_2+q)}{D_{\rho13}}-\frac{(q_4,2q-2q_2+q_3)}{2m^2_\rho}\right]
+\nonumber\\&&\frac{(q_4,q_2-q_3)}{4D_{\rho23}}
+\frac{(q_4,q_2+q_3)}{4m^2_\rho}-\nonumber\\&&
\frac{3(q,q_4)-m^2_\pi-D_\rho(q)}{4D_{a_1}(q-q_3)}
\left[\frac{(q_2,4q_3-q_4+q)}{D_{\rho13}}-\right.
\nonumber\\&&\left.\frac{(q_1,q_2-q_3)}{D_{\rho23}}-\frac{(q_3,2q-2q_4+q_2)}{2m^2_\rho}\right]
-\frac{(q_2,q_3)}{2D_{\rho14}}.\label{A1}\end{eqnarray} The
notations are:  $\widehat{P}_{ab}$ is the operator interchanging
the pion momenta $q_a\leftrightarrow q_b$,  $D_{\rho ab}\equiv
D_\rho(q_a+q_b)$ is the inverse propagator of $\rho$ meson with
the invariant mass squared $(q_a+q_b)^2$,
\begin{equation}
D_\rho(q)=m^2_\rho-q^2-i\sqrt{q^2}\Gamma_\rho(\sqrt{q^2}),\label{Drho}\end{equation}
see (3.3)$-$(3.5) in \cite{ach05} for $\Gamma_\rho(\sqrt{q^2})$.
The terms $\propto D_\rho(q)$ in (\ref{A1}) refer to the contact
terms generated by (\ref{photon}).  $(P,Q)$ stands for invariant
scalar product of two four-vectors $P$ and $Q$,
$D_\pi(p)=m^2_\pi-p^2$ is the inverse propagator of pion, $m_\pi$
and $m_\rho$ are the masses of charged pion and $\rho(770)$ meson
taken from \cite{pdg}. $A_2$ is obtained from $A_1$ by
interchanging $q_1\leftrightarrow q_2$, $A_3$ is obtained from
$A_1$ by simultaneous interchanges $q_1\leftrightarrow q_3$,
$q_2\leftrightarrow q_4$ followed by inverting an overall sign,
and $A_4$ is obtained from $A_3$ by interchanging
$q_3\leftrightarrow q_4$. The form of the $a_1$ propagator
$D^{-1}_{a_1}$ with the energy dependent width is given in
\cite{ach05}. Here  $\Gamma_{a_1}\not=0$ should be taken into
account  because $\sqrt{s}=1$ GeV is close to $m_{a_1}=1.23$ GeV
(a PDG value \cite{pdg}) or to $m_{a_1}=\sqrt{2}m_\rho=1.09$ GeV
given by  Weinberg's relation. We use the approximate expression
for $\Gamma_{a_1}(m)$ which interpolates the  curve in
\cite{ach05} in the range $3m_\pi\leq m\leq\sqrt{s}-m_\pi$,
$\sqrt{s}\leq1$ GeV.

The resonant contribution
$\gamma^\ast\to\rho\to\pi^+\pi^-\pi^+\pi^-$ in (\ref{meff})
respects the requirement of chiral symmetry in that it vanishes at
the vanishing momentum $q_{a\mu}\to 0$ ($a=1,2,3,4$) of any final
pion, provided $m_\pi=0$. However, the terms due to the direct
$\gamma^\ast\to\pi^+\pi^-\pi^+\pi^-$ contribution  do not vanish
in the above limiting cases. This is the consequence of the
breaking of conservation of the axial current by electromagnetic
field, $\partial_\mu j^a_{\mu,A}=e{\cal
A}_\mu\epsilon_{3ab}j^b_{\mu,A}$ upon neglecting the term $\propto
m_\pi^2$. One can show that the terms in (\ref{meff}) surviving in
the limit $q_{a\mu}\to 0$, correspond to the matrix elements of
the above divergence of axial current.

The results of evaluation of the $e^+e^-\to\pi^+\pi^-\pi^+\pi^-$
reaction cross section in GHLS  model are shown in Fig.~1.
\begin{figure}
\includegraphics[width=75mm]{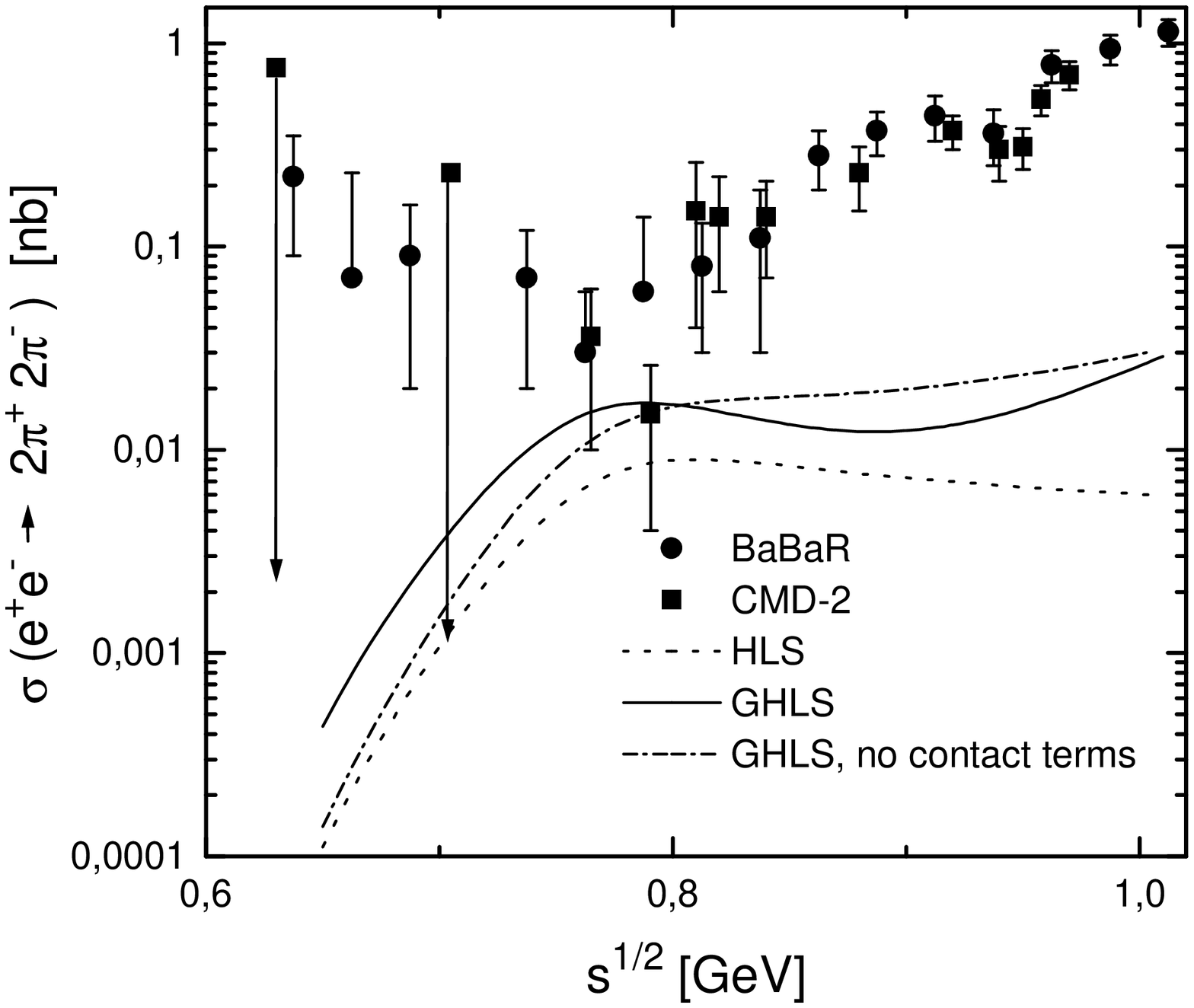}
\caption{The energy dependence of the
$e^+e^-\to\pi^+\pi^-\pi^+\pi^-$ reaction cross section in the
generalized hidden local symmetry model; $m_{a_1}=1.23$ GeV. The
data are CMD-2 \cite{cmd2} and BaBaR \cite{babar}. "HLS" refers to
the case of no $a_1$, no contact terms, "GHLS" does to one with
both  $a_1$ meson and contact couplings (\ref{photon}). "GHLS, no
contact terms" refers to the model without contact terms.}
\end{figure}
The curves are obtained in the case $m_{a_1}=1.23$ GeV; the
results for the mass $m_{a_1}=1.09$ GeV look qualitatively the
same. One can see that the model is unable to reproduce the
magnitude of the cross section at energies $\sqrt{s}>0.8$ GeV. Let
us include the contributions of heavier resonances
$\rho^\prime\equiv\rho(1450)$ and
$\rho^{\prime\prime}\equiv\rho(1700)$ trying to explain the cross
section magnitude at $\sqrt{s}\geq0.8$ GeV, without invoking the
higher derivative terms in the effective lagrangian.   We choose
the simplest parametrization consisting of the Breit-Wigner
resonance shape with the constant widths and masses
$m_{\rho^\prime}=1.459$ GeV, $\Gamma_{\rho^\prime}=0.147$ GeV,
$m_{\rho^{\prime\prime}}=1.72$ GeV,
$\Gamma_{\rho^{\prime\prime}}=0.25$ GeV taken from \cite{pdg} and
neglect the $\rho(770)-\rho(1450)-\rho(1700)$ mixing due to their
common decay modes. This approximation  results in no qualitative
difference in the role of heavy resonance at $\sqrt{s}\leq1$ GeV
as compared to more sophisticated models with mixing. We also
adopt the assumption  of $a_1\pi$ dominance in the
$\rho^\prime,\rho^{\prime\prime}\to4\pi$ decay dynamics
\cite{cmd2_99}, but modify it to include the requirements of
chiral symmetry. Then taking into account the
$\rho^\prime,\rho^{\prime\prime}$ resonance contributions results
in the factor
\begin{equation}
R(s)=\left|1+\frac{D_\rho(q)}{1+r(s)}\left[\frac{x_{\rho^\prime}}{D_{\rho^\prime}(q)}+
\frac{x_{\rho^{\prime\prime}}}{D_{\rho^{\prime\prime}}(q)}\right]\right|^2,
\label{rhoprime}
\end{equation}multiplying the right hand side of
(\ref{curve}), where $D_V(q)=m^2_V-s-im_V\Gamma_V$,
$V=\rho^\prime,\rho^{\prime\prime}$, $s=q^2$. Free parameters
$x_{\rho^\prime}$ and $x_{\rho^{\prime\prime}}$ are found  from
fitting the data. The meaning of $x_{\rho^\prime}$ is that
\begin{equation}
x_{\rho^\prime}=\frac{g_{\gamma\rho^\prime}}{g_{\gamma\rho}}\frac{g
_{\rho^\prime\to a_1\pi\to4\pi}}{g_{\rho\to
a_1\pi\to4\pi}},\label{xpr1}\end{equation} analogously for
$x_{\rho^{\prime\prime}}$, where $g_{\gamma V}=em^2_V/f_V$ is the
photon-vector meson $V$ transition amplitude,  $f_V$ is related
with the leptonic width (\ref{leptwidth}). Since $\rho$ and
$\rho^\prime$ are assumed here to have the similar coupling to the
state $a_1\pi$, the ratio (\ref{xpr1}) is constant.  The complex
function $r(s)$ in (\ref{rhoprime}) is the ratio of the amplitude
with the intermediate $a_1$ meson to one with no $a_1$
contribution. It approximately takes into account the $a_1\pi$
dominance in the four pion decay of heavy isovector resonances and
is precalculated for the CMD-2 \cite{cmd2} and BaBaR \cite{babar}
data points $\sqrt{s}\leq1$ GeV:
\begin{eqnarray}
r(s)&=&\left[\frac{\Gamma^{{\rm eff, no}a_1}_{\rho\to
4\pi}}{\Gamma_{\rho\to
a_1\pi\to4\pi}}\right]^{1/2}\exp(i\chi),\nonumber\\
\chi&=&\cos^{-1}\frac{\Gamma^{\rm eff}_{\rho\to4\pi}-\Gamma^{{\rm
eff, no}a_1}_{\rho\to 4\pi}-\Gamma_{\rho\to
a_1\pi\to4\pi}}{2\sqrt{\Gamma_{\rho\to a_1\pi\to4\pi}\Gamma^{{\rm
eff, no}a_1}_{\rho\to 4\pi}}}.\label{r}\end{eqnarray} Here
$\Gamma_{\rho\to a_1\pi\to4\pi}\equiv\Gamma_{\rho\to
a_1\pi\to4\pi}(s)$ is the $\rho^0\to\pi^+\pi^-\pi^+\pi^-$ decay
width due to the intermediate $a_1\pi$ state only, while
$\Gamma^{{\rm eff, no}a_1}_{\rho\to 4\pi}\equiv\Gamma^{{\rm eff,
no}a_1}_{\rho\to 4\pi}(s)$ is the effective width of the same
decay including all the contribution mentioned above except the
$a_1\pi$ one. The approximation (\ref{r}) corresponds to the
averaging over four pion phase space necessary to evade
unacceptably long time  in the fitting procedure.

The results of fitting the CMD-2 data are given in Table 1.
\begin{table}
\caption{\label{rescmd2}Table 1. The results of fitting CMD-2 data
\cite{cmd2}. }
\begin{tabular}{ccccc}
\hline &$x_{\rho^\prime}$
&$x_{\rho^{\prime\prime}}$&$\chi^2/N_{\rm
d.o.f}$&$m_{a_1}$ [GeV]\\
\hline1& $-27.5\pm1.5$&$\equiv0$&15.4/10&1.23\\
      2& $\equiv0$&$-46.2\pm2.5$&15.4/10&1.23\\
      3& $96.8\pm1.5$&$-208.7\pm2.5$&14.5/9&1.23\\
      4& $-17.8\pm1.0$&$\equiv0$&15.7/10&1.09\\
      5& $\equiv0$&$-30.1\pm1.5$&15.4/10&1.09\\
      6& $72.5\pm1.0$&$-151.9\pm1.6$&14.7/9&1.09\\ \hline
\end{tabular}
\end{table}
The curves corresponding to the fit variant 3 with $\rho^\prime$
and $\rho^{\prime\prime}$ resonances  are shown in Fig.~2.
\begin{figure}
\includegraphics[width=75mm]{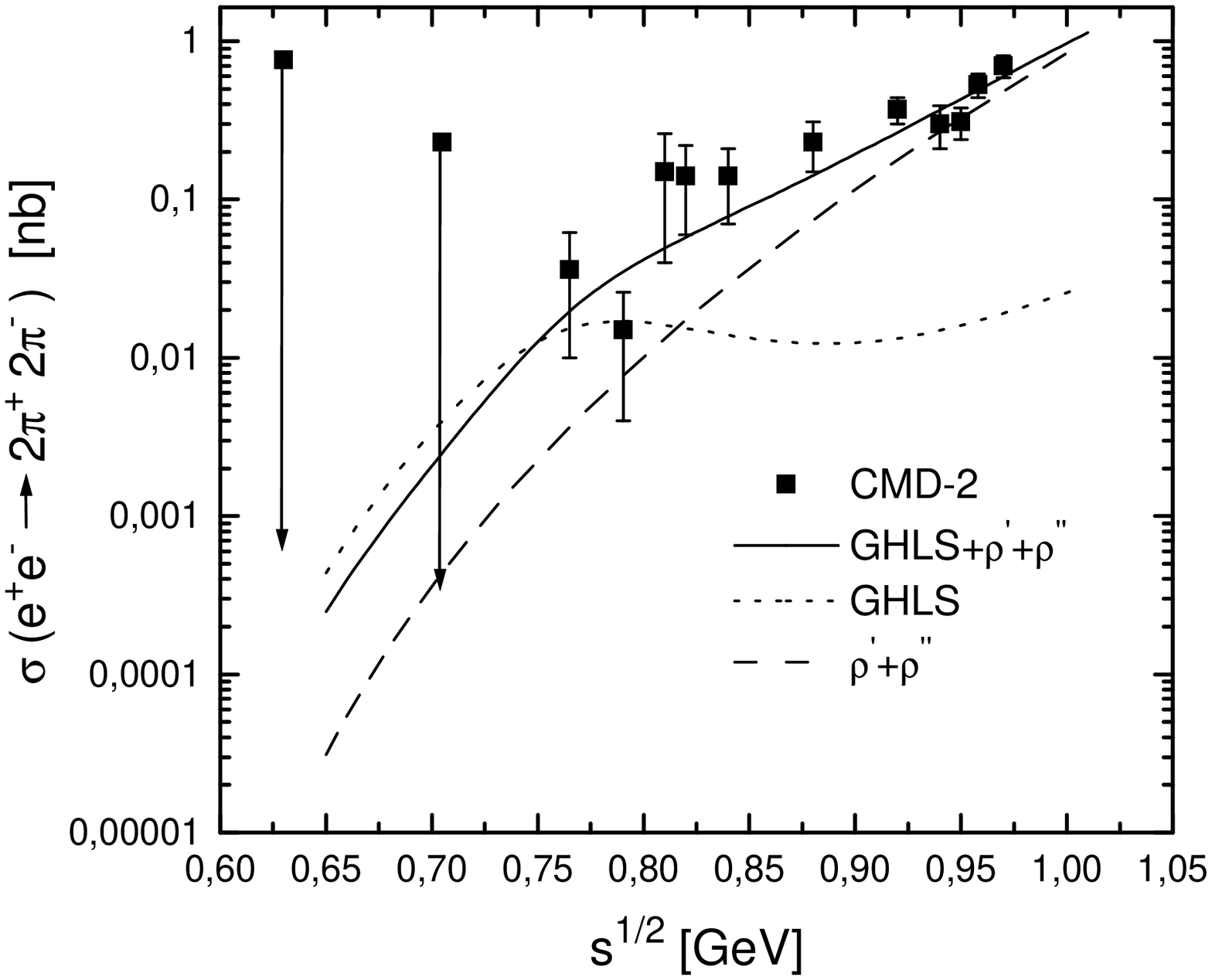}
\caption{The results of fitting the CMD-2 data \cite{cmd2}. "GHLS"
refers to the model without $\rho^\prime$, $\rho^{\prime\prime}$.}
\end{figure}
This variant is indistinguishable from the variants with the
single $\rho^\prime$ (variant 1) or $\rho^{\prime\prime}$ (variant
2), both resulting in the same curves as the dashed one shown in
Fig.~2.  Variants $4-6$ correspond to the fits with the mass
$m_{a_1}=m_\rho\sqrt{2}=1.09$ GeV and result in the same
corresponding curves not shown here. The quality of fit is not
quite good. Nevertheless, we quote the contribution of the sum
$\rho^\prime+\rho^{\prime\prime}$ (variant 3) or $\rho^\prime$
(variant 1) and $\rho^{\prime\prime}$ (variant 2) relative to the
case of pure GHLS contribution (dotted line in Fig.~2) to be 0.3
at $\sqrt{s}\approx m_\rho$ and 32 at $\sqrt{s}=1$ GeV. These
numbers refer to the case $m_{a_1}=1.23$ GeV. The case
$m_{a_1}=1.09$ GeV results in almost the same figures for above
ratios.

The results of the similar analysis of the BaBaR data \cite{babar}
are presented in Table 2.
\begin{table}
\caption{\label{resbabar}Table 2. The results of fitting BaBaR
data \cite{babar}. }
\begin{tabular}{ccccc}\hline
&$x_{\rho^\prime}$ &$x_{\rho^{\prime\prime}}$&$\chi^2/N_{\rm
d.o.f}$&$m_{a_1}$ [GeV]\\
\hline1& $-25.2\pm0.9$&$\equiv0$&32.6/16&1.23\\
      2& $\equiv0$&$-44.0\pm2.1$&29.3/16&1.23\\
      3& $273.2\pm1.4$&$-514.5\pm2.3$&11.2/15&1.23\\
      4& $-15.8\pm0.8$&$\equiv0$&35.0/16&1.09\\
      5& $\equiv0$&$-27.7\pm1.3$&31.8/16&1.09\\
      6& $198.5\pm1.0$&$-370.1\pm1.5$&11.2/15&1.09\\\hline
\end{tabular}
\end{table}
Contrary to the previous case, here the variants with the single
additional heavy resonance give a bad description. The fit chooses
two destructively interfering $\rho^\prime$ and
$\rho^{\prime\prime}$ resonances each coupled to $a_1\pi$ much
strongly than in the variants of the single heavy resonance. The
curves shown in Fig.~3 refer to variant 3 in Table 2 with
$m_{a_1}=1.23$ GeV.
\begin{figure}
\includegraphics[width=75mm]{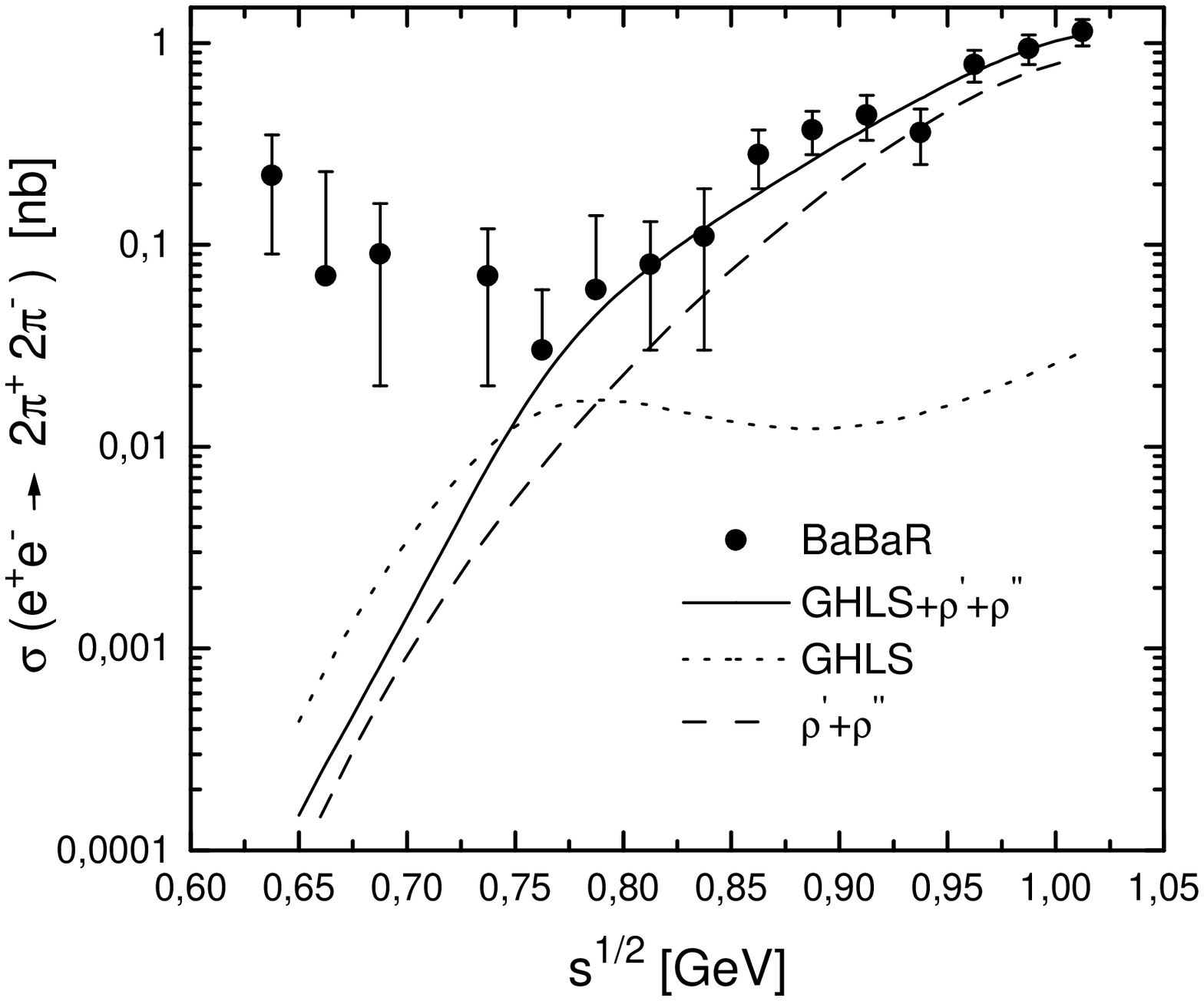}
\caption{The same as in Fig.~2, but for the BaBaR data
\cite{babar}.}
\end{figure}
The contribution of the sum $\rho^\prime+\rho^{\prime\prime}$ ( in
variant 3) or $\rho^\prime$ (variant 1) and $\rho^{\prime\prime}$
(variant 2) relative to the case of pure GHLS contribution (dotted
line in Fig.~3)is found  to be 0.6 at $\sqrt{s}\approx m_\rho$ and
30 at $\sqrt{s}=1$ GeV. As in the case of the CMD-2 data, here the
variant 6 with $m_{a_1}=1.09$ GeV results in practically the same
corresponding curves and ratios.

Our conclusions differ from the result of the works
\cite{cmd2,ecker,lichard} all claiming small or even absent
contribution of heavy resonances.  We attribute this disagreement
to the difference among the models used in the present  analysis
and in works \cite{cmd2,cmd2_99,ecker,lichard}. The works
\cite{cmd2,lichard} exploit non-chiral invariant effective
Lagrangians. The work \cite{ecker} is based on chiral amplitude
with three unknown parameters. No central values nor their errors
are given in order to assess independently the quality of approach
\cite{ecker}. The effective vertex $a_1\rho\pi$ used in that work
refers to the higher derivative contribution, while there exists a
lowest derivative one used in the present work, see \cite{ach05}.
The contact $\gamma\pi^+\pi^-$ vertex is present in the
intermediate state of the amplitude in \cite{ecker}. The apparent
violation of the vector dominance of the pion form factor could be
evaded  by adjusting arbitrary constants in  in \cite{ecker} only
assuming the vanishing of the $\rho$ meson width which is
inappropriate in the energy range where the $\rho$ width is
essential.

Thus, the simplest variant of GHLS model  with the minimal number
of derivatives fails to explain the cross section of the reaction
$e^+e^-\to\pi^+\pi^-\pi^+\pi^-$ at energies $0.8<\sqrt{s}\leq1$
GeV. One possible way out this difficulty by including heavy
resonances $\rho^\prime$, $\rho^{\prime\prime}$ is studied here.
GHLS  model  is based on the nonlinear realization of chiral
symmetry. It would be desirable to readdress the present issues in
the frame work of the chiral model of the vector and axial vector
mesons based on the linear $\sigma$-model. This task is necessary
in order to evaluate the robustness of the figures characterizing
the contributions of heavier resonances towards various model
assumptions and to reveal the role of the intermediate states
which include the widely discussed scalar $\sigma$ meson.

The work is partially supported by the grants of the Russian
Foundation for Basic Research  RFBR-07-02-00093 and of the Support
of the Leading Scientific Schools NSh-5362.2006.2.


\begin{thebibliography}{99}
\bibitem{meissner88}
Ulf-G.~Meissner, Phys.Rept. {\bf161}, 213 (1988).

\bibitem{bando88}
M.~Bando, T.~Kugo, and  K.~Yamawaki, Phys.Rept. {\bf164}, 217
(1988).


\bibitem{cmd2}
R.~R.~Akhmetshin {\it et al.} (CMD-2 Collab.), Phys.Lett.
B{\bf475}, 190 (2000) [arXiv:hep-ex/9912020v1].

\bibitem{babar}
B.~Aubert {\it et al.} (BaBaR Collab.), Phys.Rev. D{\bf71}, 052001
(2005) [arXiv:hep-ex/0502025v1].



\bibitem{ach00b}
N.~N.~Achasov and A.~A.~Kozhevnikov, Phys.Rev. D{\bf62}, 056011
(2000) [arXiv:hep-ph/0003094v3];  Zh.Eksp.Teor.Fiz. {\bf 91}, 499
(2000).

\bibitem{ach05}
N.~N.~Achasov and A.~A.~Kozhevnikov, Phys.Rev. D{\bf71}, 034015
(2005) [arXiv:hep-ph/0412077v2]; Yad.Fiz. {\bf69}, 314 (2006).

\bibitem{pdg}
W.~M.~Yao {\it et al.} (Particle Data Group), J.Phys.G:
Nucl.Part.Phys. {\bf33}, 1 (2006).



\bibitem{cmd2_99}
R.~R.~Akhmetshin {\it et al.} (CMD-2 Collab.), Phys.Lett. B
{\bf466}, 392 (1999) [arXiv:hep-ex/9904024v2].

\bibitem{ecker}
G.~Ecker and R.~Unterdorfer, Eur.Phys.J. C{\bf24}, 535 (2002)
[arXiv:hep-ph/0203075v2].

\bibitem{lichard}
P.~Lichard and J.~Jur$\acute{\rm a}\check{\rm n}$, Phys.Rev. D{\bf
76}, 094030 (2007) [arXiv:hep-ph/0601234v2].
\end{thebibliography}
\end{document}